\documentclass[11pt,english,twocolumn]{smfart}
\usepackage{mathptmx}
\title{The supersymmetry method of random matrix theory}
\author{Martin R. Zirnbauer}
\address{Institut f\"ur Theoretische Physik,
Uni\-versit\"at zu K\"oln, Z\"ulpicher Str.\ 77, 50937 K\"oln,
Germany}
\date{December 31, 2003}
\email{zirn@thp.Uni-Koeln.DE}
\begin{document}
\maketitle

\section{Introduction}

A prominent theme of modern condensed matter phy\-sics is
electronic transport -- in particular the electrical conductivity
-- of disordered metallic systems at very low temperatures. From
the Landau theory of weakly interacting Fermi liquids one expects
the essential aspects of the situation to be captured by the
single-elec\-tron approximation. Mathematical models that have
been proposed and studied in this context include random
Schr\"odinger operators and band random matrices.

If the physical system has infinite size, two distinct
possibilities exist: the quantum single-electron motion may either
be bounded or unbounded.  In the former case the disordered
electron system is an insulator, in the latter case a metal with
finite conductivity (if the electron motion is not critical but
diffusive). Metallic behavior is expected for weakly disordered
systems in three dimensions; insulating behavior sets in when the
disorder is increased or the space dimension reduced.

The main theoretical tool used in the physics literature on the
subject is the {\it supersymmetry method} pioneered by Wegner and
Efetov (1979-1983).  Over the past twenty years, physicists have
applied the method in many instances, and a rather complete
picture of weakly disordered metals has emerged. Several excellent
reviews of these developments are available in print.

From the perspective of mathematics, however, the method has not
always been described correctly, and what is sorely lacking at
present is an exposition of how to implement the method {\it
rigorously}. [Unfortunately, the correct exposition by Sch\"afer
and Wegner (1980) was largely ignored or forgotten by later
authors.]  In this encyclopedia article an attempt will be made to
help remedy the situation, by giving a careful review of the
Wegner-Efetov supersymmetry method for the case of Hermitian band
random matrices.

\section{Gaussian Ensembles}

Let $V$ be a unitary vector space of finite dimension.  A
Hermitian random matrix model on $V$ is defined by some
probability distribution on ${\rm Herm}(V)$, the Hermitian linear
operators on $V$. You may fix some orthonormal basis of $V$ and
represent the elements $H$ of ${\rm Herm}(V)$ by Hermitian square
matrices.

Quite generally, probability distributions are characterized by
their Fourier transform or characteristic function.  In the
present case this is
\begin{displaymath}
  \Omega(K) = \big\langle {\rm e}^{ {\rm i} {\rm Tr}\,  H K}
  \big\rangle \;,
\end{displaymath}
where the Fourier variable $K$ is some other linear operator on
$V$, and $\langle \ldots \rangle$ denotes the expectation value
w.r.t.~the probability distribution for $H$. Later it will be
important that, if $\Omega(K)$ is an analytic function of $K$, the
matrix entries of $K$ need not be from $\mathbb{R}$ or
$\mathbb{C}$ but can be taken from the even part of some exterior
algebra.

The probability distributions to be considered in this article are
Gaussian with zero mean, $\langle H \rangle = 0$.  Their Fourier
transform is also Gaussian:
\begin{displaymath}
  \Omega(K) = {\rm e}^{- \frac{1}{2} J(K,K)} \;,
\end{displaymath}
with $J$ some quadratic form.  We now describe $J$ for a large
family of hierarchical models that includes the case of band
random matrices.

Let $V$ be given a decomposition by orthogonal vector spaces:
\begin{displaymath}
  V = V_1 \oplus V_2 \oplus \ldots \oplus V_{|\Lambda|} \;.
\end{displaymath}
You should imagine that every vector space $V_i$ corresponds to
one site $i$ of some lattice $\Lambda$, and the total number of
sites is $|\Lambda|$.  For simplicity, we take all dimensions to
be equal: ${\rm dim}\, V_1 = \ldots = {\rm dim}\, V_{|\Lambda|} =
N$. Thus the dimension of $V$ is $N |\Lambda|$.  The integer $N$
is called the number of orbitals per site.

If $\Pi_i$ is the orthogonal projector on the linear subspace $V_i
\subset V$, we take the bilinear form $J$ to be
\begin{displaymath}
  J(K,K^\prime) = \sum_{i,j = 1}^{|\Lambda|} J_{ij} \,
  {\rm Tr}(\Pi_i \, K \, \Pi_j \, K^\prime ) \;,
\end{displaymath}
where the coefficients $J_{ij}$ are real, symmetric, and positive.
This choice of $J$ implies invariance under the group ${\mathcal
U}$ of unitary transformations in each subspace:
\begin{displaymath}
  {\mathcal U} = {\rm U}(V_1) \times {\rm U}(V_2) \times \cdots
  \times {\rm U}(V_{|\Lambda|}) \;.
\end{displaymath}
Clearly, $\Omega(K) = \Omega(U K U^{-1})$ or, equivalently, the
probability distribution for $H$ is invariant under conjugation $H
\mapsto U H U^{-1}$, for $U \in {\mathcal U}$.

If $\{ e_i^a \}_{a = 1, \ldots, N}$ is an orthonormal basis of $V_i$,
we define linear operators $E_{ij}^{ab} : V_j \to V_i$ by $E_{ij}^{ab}
e_j^b = e_i^a$.  By evaluating $J(E_{ij}^{ab}, E_{j^ \prime
i^\prime}^{b^\prime a^\prime}) = J_{ij} \delta_{i i^\prime} \delta_{j
j^\prime} \delta^{a a^\prime} \delta^{b b^\prime}$ one sees that the
matrix entries of $H$ are all statistically independent.

By varying the lattice $\Lambda$, the number of orbitals $N$, and
the variances $J_{ij}$, one obtains a large class of Hermitian
random matrix models, two prominent subclasses of which are the
following:
\begin{enumerate}
\item For $|\Lambda| = 1$, one gets the Gaussian Unitary Ensemble
(GUE).  Its symmetry group is ${\mathcal U} = {\rm U}(N)$, the
largest one possible in dimension $N = {\rm dim}V$.
\item If $|i - j|$ denotes a distance function for $\Lambda$, and
$f$ a rapidly decreasing positive function on $\mathbb{R}_+$ of
width $W$, the choice $J_{ij} = f(|i-j|)$ with $N = 1$ gives an
ensemble of band random matrices with band width $W$ and symmetry
group ${\mathcal U} = {\rm U}(1)^{|\Lambda|}$.
\end{enumerate}
Beyond being real, symmetric and positive, the variances $J_{ij}$
are required to have two extra properties in order for all of the
following treatment to go through:
\begin{itemize}
{\item[$\bullet$] They must be positive as a quadratic form. This
is to guarantee the existence of an inverse, which we denote by
$w_{ij} = (J^{-1})_{ij}$.}
{\item[$\bullet$] The off-diagonal matrix entries of the inverse
must be non-positive: $w_{ij} \le 0$ for $i \not= j$.}
\end{itemize}

\section{Basic tools}

\subsection{Green's functions.}\label{sec:green}

A major goal of random-matrix theory is to understand the
statistical behavior of the spectrum and the eigenstates of a
random Hamiltonian $H$.  Spectral and eigenstate information can
be extracted from the Green's function, i.e.~from matrix elements
of the operator $(z - H)^{-1}$ with complex parameter $z \in
\mathbb{C} \setminus \mathbb{R}$. For the models at hand, the good
objects to consider are averages of ${\mathcal U} $--invariant
observables such as
\begin{eqnarray}
    &&G_{i}^{(1)}(z) = \big\langle {\rm Tr} \, \Pi_i
    (z - H)^{-1} \big\rangle \;, \label{onepoint} \\
    &&G_{ij}^{(2)}(z_1,z_2) = \big\langle {\rm Tr} \, \Pi_i
    (z_1 - H)^{-1} \Pi_j (z_2 - H)^{-1} \big\rangle
    \label{twopoint} \;.
\end{eqnarray}
The discontinuity of $G_i^{(1)}(z)$ across the real $z$--axis
yields the local density of states.  In the limit of infinite
volume $(|\Lambda| \to \infty$), the function $G_{ij}^{(2)}
(z_1,z_2)$ for $z_1 = E + {\rm i}\epsilon$, $z_2 = E - {\rm
i}\epsilon$, real energy $E$, and $\epsilon > 0$ going zero, gives
information on transport, e.g.~the electrical conductivity by the
Kubo-Greenwood formula.

Mathematically speaking, if $G_{ij}^{(2)} (E + {\rm i} \epsilon, E
- {\rm i} \epsilon)$ is bounded (for infinite volume) in
$\epsilon$ and decays algebraically with distance $|i - j|$ at
$\epsilon = 0+$, the spectrum is absolutely continuous and the
eigenstates are extended at energy $E$. On the other hand, a pure
point spectrum and localized eigenstates are signalled by the
behavior $G_{ij}^{(2)} \sim \epsilon^{-1} {\rm e}^{-\lambda |i -
j|}$ with positive Lyapunov exponent $\lambda$.

\subsection{Green's functions from determinants.}\label{sec:gfd}

For any pair of linear operators $A, B$ on a finite-dimensio\-nal
vector space $V$, the following formula from basic linear algebra
holds if $A$ has an inverse:
\begin{displaymath}
    \frac{d}{dt} {\rm Det}(A + tB) \Big|_{t = 0} =
    {\rm Det}(A) \, {\rm Tr} (A^{-1} B) \;.
\end{displaymath}
Using it with $A = z - H$ and $z \in \mathbb{C} \setminus
\mathbb{R}$, all Green's functions can be expressed in terms of
determinants; for example, $G_{ij}^{(2)}(w,z) =$
\begin{displaymath}
    \sum_{a,b = 1}^N \frac{\partial^2}{\partial s \partial t}
    \left\langle \frac{{\rm Det}(w - H) {\rm Det}(z - H + t
    E_{ij}^{ab})} {{\rm Det}(w - H - s E_{ji}^{ba}) {\rm Det}(z-H)}
    \right\rangle \Bigg|_{s = t = 0} .
\end{displaymath}
It is clear that, given a formula of this kind, what one wants is
a method to handle ensemble averages of ratios of determinants.
This is what's reviewed in the sequel.

\subsection{Determinants as Gaussian integrals.}\label{sec:dgi}

Let the Hermitian scalar product of the unitary vector space $V$
be written as $\varphi_1, \varphi_2 \mapsto (\bar\varphi_1 ,
\varphi_2)$, and denote the adjoint or Hermitian conjugate of a
linear operator $A$ on $V$ by $A^\ast$.  If $\mathfrak{Re}\, A :=
\frac{1}{2}(A + A^\ast) > 0$, the standard Lebesgue integral of
the Gaussian function $\varphi \mapsto {\rm e}^{- (\bar\varphi, A
\varphi)}$ makes sense and gives
\begin{equation}\label{bosgauss}
  \int {\rm e}^{ - (\bar\varphi, A \varphi)} = {\rm Det}\,
  A^{-1} \;,
\end{equation}
where it is understood that we are integrating with the Lebesgue
measure on (the normed vector space) $V$ normalized by $\int {\rm
e}^{- (\bar\varphi , \varphi)} = 1$.  The same integral with
\emph{anticommuting} $\psi$ instead of the (commuting) $\varphi
\in V$ gives
\begin{equation}\label{fermgauss}
    \int {\rm e}^{ - (\bar\psi, A \psi) } = {\rm Det} \, A \;.
\end{equation}
This basic formula from the field theory of fermionic particles is
a consequence of the integration over anti-commuting variables
actually being \emph{differentiation}:
\begin{displaymath}
\int d\bar\psi_1 d\psi_1 f(\bar\psi_1, \psi_1, \ldots) :=
\frac{\partial^2}{\partial \bar\psi_1 \partial \psi_1}
f(\bar\psi_1, \psi_1, \ldots) \;.
\end{displaymath}

\section{Fermionic variant}\label{sec:ferm}

The supersymmetry method of random-matrix theory is a theme with
many variations. The first variation to be described is the
``fermionic'' one. To optimize the notation, we now write
$d\mu_{N,J}(H)$ for the density of the Gaussian probability
distribution of $H$:

\begin{displaymath}
\langle F(H) \rangle = \int F(H) \, d\mu_{N,J}(H) \;.
\end{displaymath}
All determinants and traces appearing below will be taken over
vector spaces that are clear from the context.

Let $z_1, \ldots, z_n$ be any set of $n$ complex numbers, put $z
:= {\rm diag}(z_1, \ldots, z_n)$ for later purposes, and consider
\begin{equation}\label{ferm_omeg_H}
    \Omega_{n,N}^{\rm ferm}(z,J) = \int \prod_{\alpha = 1}^n
    {\rm Det} (z_\alpha - H) \, d\mu_{N,J}(H) \;.
\end{equation}
The supersymmetry method expresses this average of a product of
determinants in an alternative way,  by integrating over a
``dual'' measure as follows.

Introducing an auxiliary unitary vector space $\mathbb{C}^n$, one
associates with every site $i$ of the lattice $\Lambda$ an object
$Q_i \in {\rm Herm}(\mathbb{C}^n)$, the space of Hermitian $n
\times n$ matrices. If $dQ_i$ for $i = 1, \ldots, |\Lambda|$ are
Lebesgue measures on ${\rm Herm}(\mathbb{C}^n)$, one puts $DQ =
{\rm const} \times \prod_i dQ_i$ and
\begin{equation}\label{measure}
    d\nu_{n,J}(Q) := {\rm e}^{- \frac{1}{2} \sum_{i,j}
    (J^{-1})_{ij} \, {\rm Tr} \, Q_i Q_j} DQ \;.
\end{equation}
The multiplicative constant in $DQ$ is fixed by requiring the
density to be normalized: $\int d\nu_{n,J}(Q) = 1$. By completing
the square, this Gaussian probability measure has the
characteristic function
\begin{displaymath}
    \int {\rm e}^{{\rm i} \sum_j {\rm Tr} \, Q_j K_j}
    \, d\nu_{n,J}(Q) = {\rm e}^{- \frac{1}{2} \sum_{ij} J_{ij}
    \, {\rm Tr} \, K_i K_j} \;,
\end{displaymath}
where the Fourier variables $K_1, \ldots, K_{|\Lambda|}$ are $n
\times n$ matrices with matrix entries taken from $\mathbb{C}$ or
another commutative algebra.

The key relation of the fermionic variant of the supersymmetry method
is that the expectation of the product of determinants
(\ref{ferm_omeg_H}) has another expression as
\begin{equation}\label{ferm_omeg_Q}
    \Omega_{n,N}^{\rm ferm}(z,J) = \int \prod_{j = 1}^{|\Lambda|}
    {\rm Det}^N (z - {\rm i} Q_j) \, d\nu_{n,J}(Q)
\end{equation}
(${\rm i} = \sqrt{-1}$). The strategy of the proof is quite
simple: one writes the determinants in both expressions for
$\Omega_{n,N}^{\rm ferm}$ as Gaussian integrals over $n N
|\Lambda|$ complex fermio\-nic variables $\psi_1, \ldots, \psi_n$
(each $\psi_\alpha$ is a vector in $V$ with anti-commuting
coefficients), using the basic formula (\ref{fermgauss}).  The
integrals then encountered are essentially the Fourier transforms
of the distributions $d\mu_{N,J}(H)$ resp. $d\nu_{n,J}(Q)$. The
result is
\begin{displaymath}
    \int {\rm e}^{- \sum_\gamma z_\gamma (\bar\psi_\gamma ,
    \psi_\gamma)} \, {\rm e}^{ - \frac{1}{2} \sum_{ij} J_{ij}
    \sum_{\alpha\beta} (\bar\psi_\alpha, \Pi_i \psi_\beta)
    (\bar\psi_\beta, \Pi_j \psi_\alpha)}
\end{displaymath}
for both expressions of $\Omega_{n,N}^{\rm ferm}$.  In other
words, although the probability distributions $d\mu_{N,J}(H)$ and
$d\nu_{n,J}(Q)$ are distinct (they are defined on different
spaces), their characteristic functions coincide when evaluated on
the Fou\-rier variables $K = \sum_\alpha \psi_\alpha (\bar
\psi_\alpha , \bullet)$ for $H$ and $(K_i)_{ \alpha\beta} =
(\bar\psi_\alpha, \Pi_i \psi_\beta)$ for $Q_i$. This estab\-lishes
the claimed equality of the expressions (\ref{ferm_omeg_H}) and
(\ref{ferm_omeg_Q}) for $\Omega_{n,N}^{\rm ferm}(z,J)$.

What's the advantage of passing to the alternative expression by
$d\nu_{n,J}(Q)$?  The answer is that, while $H$ is made up of
\emph{independent} random variables, the new variables $Q_i$,
called the {\it Hubbard-Stratonovich field}, are
\emph{correlated}: they interact through the ``exchange''
constants $w_{ij} = (J^{-1})_{ij}$.  If that interaction creates
enough collectivity, a kind of mean-field behavior results.

For the simple case of GUE ($|\Lambda| = 1$, $w_{11} = N /
\lambda^2$) with $z_1 = \ldots = z_n = E$, one gets the relation
\begin{displaymath}
    \big\langle {\rm Det}^n (E - H) \big\rangle = \int
    {\rm Det}^N (E - {\rm i}Q)\,  {\rm e}^{-\frac{N}{2\lambda^2}
    {\rm Tr} \, Q^2} dQ \;,
\end{displaymath}
the right-hand side of which is easily analyzed by the steepest
descent method in the limit of large $N$.

For band random matrices in the so-called ergodic regime the
physical behavior turns out to be governed by the constant mode
$Q_1 = \ldots = Q_{|\Lambda|}$ -- a fact that can be used to
establish GUE universality in that regime.

\section{Bosonic variant}\label{sec:bos}

The bosonic variant of the present method, due to Wegner, computes
averages of products of determinants placed in the denominator:
\begin{equation}\label{invdet_H}
    \Omega_{n,N}^{\rm bos}(z,J) = \int \prod_{\alpha = 1}^n
    {\rm Det}^{-1} (z_\alpha - H) \, d\mu_{N,J}(H) \;,
\end{equation}
where we now require $\mathfrak{Im} \, z_\alpha \not= 0$ for all
$\alpha = 1, \ldots, n$. Complications relative to the fermionic
case arise from the fact that the integrand in (\ref{invdet_H})
has poles.  If one replaces the anti-commuting vectors
$\psi_\alpha$ by commuting ones $\varphi_\alpha$, and then simply
repeats the previous calculation in a naive manner, one arrives at
\begin{equation}\label{naive}
    \Omega_{n,N}^{\rm bos}(z,J) \stackrel{?}{=}
    \int \prod_{j = 1}^{|\Lambda|}
    {\rm Det}^{-N} (z - Q_j) \, d\nu_{n,J}(Q) \;,
\end{equation}
where the integral is still over $Q_j \in {\rm Herm} (\mathbb{
C}^n)$. The calculation is correct, and relation (\ref{naive})
therefore holds true, provided that the parameters $z_1, \ldots,
z_n$ all lie in the \emph{same half} (upper or lower) of the
complex plane. To obtain information on transport properties,
however, one needs parameters in both the upper and lower halves;
see the paragraph following (\ref{twopoint}). The general case to
be addressed below is $\mathfrak{Im} \, z_\alpha > 0$ for $\alpha
= 1, \ldots, p$, and $\mathfrak{Im} \, z_\alpha < 0$ for $\alpha =
p+1, \ldots, n$. Careful inspection of the steps leading to
equation (\ref{naive}) reveals a convergence problem for $0 < p <
n$. In fact, (\ref{naive}) with $Q_j$ in ${\rm Herm}(\mathbb{
C}^n)$ turns out to be \emph{false} in that range. Learning how to
resolve this problem is the main step toward mathematical mastery
of the method. Let us therefore give the details.

If $s_\alpha := {\rm sgn} \, \mathfrak{Im} \, z_\alpha$, the good
(meaning convergent) Gaussian integral to consider is
\begin{displaymath}
    \int {\rm e}^{{\rm i} \sum_\alpha s_\alpha \big( \bar\varphi_\alpha ,
    (z_\alpha - H) \varphi_\alpha \big)} = \prod_{\alpha = 1}^n {\rm
    Det}^{-1} \big( - {\rm i}s_\alpha (z_\alpha - H) \big) \;.
\end{displaymath}
To avoid carrying around trivial constants, we now assume ${\rm
  i}^{(n - 2p) N |\Lambda|} = 1$.  Use of the characteristic function
of the distribution for $H$ then gives
\begin{eqnarray}
    &&\Omega_{n,N}^{\rm bos}(z,J) = \int {\rm e}^{{\rm i} \sum_\gamma
    s_\gamma z_\gamma (\bar\varphi_\gamma, \varphi_\gamma) }
    \label{difficult} \\ &&\times \, {\rm e}^{-
    \frac{1}{2} \sum_{ij} J_{ij} \sum_{\alpha\beta} s_\alpha
    (\bar\varphi_\alpha , \Pi_i \varphi_\beta) s_\beta
    (\bar\varphi_\beta , \Pi_j \varphi_\alpha)} \nonumber \;.
\end{eqnarray}
The difficulty of analyzing this expression stems from the
``hyperbolic'' nature (due to the indefiniteness of the signs
$s_\alpha = \pm 1$) of the term quartic in the $\varphi_\alpha,
\bar\varphi_\alpha$.

\subsection{Fyodorov's method}

The integrand for $\Omega^{\rm bos}$ is naturally expressed in
terms of $n \times n$ matrices $M_i$ with matrix elements
$(M_i)_{\alpha\beta} = (\bar\varphi_\alpha, \Pi_i \varphi_\beta)$.
These matrices lie in ${\rm Herm}^+ (\mathbb{ C}^n)$, i.e.~they
are non-negative as well as Hermitian. Fyodorov's idea was to
introduce them as the new variables of integration.  To do that
step recall the basic fact that, given two differentiable spaces
$X$ and $Y$ and a smooth map $\psi : X \to Y$, a distribution
$\mu$ on $X$ is \emph{pushed forward} to a distribution
$\psi(\mu)$ on $Y$ by $\psi(\mu)[f] := \mu[f \circ \psi]$, where
$f$ is any test function on $Y$.

We apply this universal principle to the case at hand by
identifying $X$ with $V^n$, and $Y$ with $\left( {\rm Herm}^+
(\mathbb{C}^n) \right)^{|\Lambda|}$, and $\psi$ with the mapping
that sends
\begin{displaymath}
(\varphi_1, \ldots, \varphi_n) \in X \quad \mbox{to} \quad (M_1,
\ldots, M_{|\Lambda|}) \in Y
\end{displaymath}
by $(M_i)_{\alpha\beta} = (\bar \varphi_\alpha , \Pi_i
\varphi_\beta)$.  On $X = V^n$ we are integrating with the product
Lebesgue measure normalized by $\int {\rm e}^{- \sum_\alpha (\bar
\varphi_\alpha , \varphi_\alpha)} = 1$. We now want the push
forward of this flat measure (or distribution) by the mapping
$\psi$. In general, the push forward of a measure is not
guaranteed to have a density but may be singular (like a Dirac
$\delta$-distribution). This is in fact what happens if $N < n$.
The matrices $M_i$ then have less than the maximal rank, so they
fail to be positive but possess \emph{zero eigenvalues}, which
implies that the flat measure on $X$ is pushed forward by $\psi$
into the \emph{boundary} of $Y$. For $N \ge n$, on the other hand,
the push forward measure \emph{does} have a density on $Y$; and
that density is $\prod_{i = 1}^{|\Lambda|} \left( {\rm Det}\, M_i
\right)^{N - n} dM_i$, as is seen by transforming to the
eigenvalue representation and comparing Jacobians.  The $dM_i$ are
Lebesgue measures on ${\rm Herm} (\mathbb{C}^n)$, normalized by
the condition
\begin{displaymath}
    \int_{M_i > 0} {\rm e}^{-{\rm Tr}\, M_i} ({\rm Det}\, M_i)^{N -
    n} dM_i = \int {\rm e}^{- \sum_\alpha (\bar\varphi_\alpha, \Pi_i
\varphi_\alpha)} = 1 \;.
\end{displaymath}
Assembling the sign information for $\mathfrak{Im}\, z_\alpha$ in
a diagonal matrix $s := {\rm diag}(s_1, \ldots, s_n)$, and pushing
the integral over $X$ forward to an integral over $Y$ with measure
$DM := \prod_i dM_i$, we obtain Fyodorov's formula:
\begin{eqnarray}
    &&\Omega_{n,N}^{\rm bos}(z,J) = \int_Y {\rm e}^{-\frac{1}{2}
    \sum_{ij} J_{ij} {\rm Tr} \, ( s M_i s M_j )} \label{fyodorov}\\
    &&\hspace{2cm} \times
    \, {\rm e}^{\sum_k {\rm Tr}\, \left( {\rm i} s \, z M_k
    + (N - n) \ln M_k \right)} DM \;. \nonumber
\end{eqnarray}
This formula has a number of attractive features. One is ease of
derivation, another is ready general\-iz\-ability to the case of
non-Gaussian distributions. The main disadvantage of the formula
is that it does not apply to the case of band random matrices
(because of the restriction $N \ge n$); nor does it combine nicely
with the fermionic formula (\ref{ferm_omeg_Q}) to give a
supersymmetric formalism, as one formula is built on $J_{ij}$ and
the other on $w_{ij}$.

Note that (\ref{fyodorov}) clearly displays the dependence on the
signature of $\mathfrak{Im}\, z$ : you cannot remove the $s_1,
\ldots, s_n$ from the integrand without changing the domain of
integration $Y = \left( {\rm Herm}^+ (\mathbb{C}^n) \right)^{
|\Lambda|}$.  This important feature is missing from the naive
formula (\ref{naive}).

Setting $q = n - p$, let ${\rm U}(p,q)$ be the pseudo-unitary
group of complex $n \times n$ matrices $T$ with inverse $T^{-1} =
s T^\ast s$. Since $|{\rm Det}T| = 1$ for $T \in {\rm U}(p,q)$,
the integration domain $Y$ and density $DM = \prod_i dM_i$ of
Fyodorov's formula are invariant under ${\rm U}(p,q)$
transformations $M_i \mapsto T M_i T^\ast$, and so is actually the
integrand in the limit where all parameters $z_1, \ldots, z_n$
become equal. Thus the elements of ${\rm U}(p,q)$ are global
symmetries in that limit.  This observation holds the key to
another method of transforming the expression (\ref{difficult}).

\subsection{The method of Sch\"afer and Wegner}

To rescue the naive formula (\ref{naive}), what needs to be
abandoned is the integration domain ${\rm Herm}(\mathbb{C}^n)$ for
the matrices $Q_i$.  The good domain to use was constructed by
Sch\"afer and Wegner, but was largely forgotten in later physics
work.

Writing $(M_k)_{\alpha\beta} = (\bar\varphi_ \alpha, \Pi_k
\varphi_\beta)$ as before, consider the function
\begin{equation}\label{funcFMQ}
    F_M(Q) = {\rm e}^{\frac{1}{2} \sum_{ij} w_{ij} {\rm Tr}
    \, (s Q_i + {\rm i}z) (s Q_j + {\rm i}z) - \sum_{k} {\rm Tr}\,
    M_k Q_k} \;,
\end{equation}
viewed as a \emph{holomorphic} function of
\begin{displaymath}
    Q = (Q_1, \ldots, Q_{|\Lambda|}) \in {\rm
    End}(\mathbb{C}^n)^{|\Lambda|} \;.
\end{displaymath}
If the Gaussian integral $\int F_M(Q) DQ$ with holomorphic density
$DQ = \prod_i dQ_i$ is formally carried out by completing the
square, one gets the integrand of (\ref{difficult}).  This is just
what we want, as it would allow us to pass to a $Q$-matrix
formulation akin to the one of Section \ref{sec:ferm}.  But how
can that formal step be made rigorous? To that end, one needs to
(i) construct a domain on which $|F_M(Q)|$ decays rapidly so that
the integral exists, and (ii) justify completion of the square and
shifting of variables.

To begin, take the absolute value of $F_M(Q)$.  Putting $\frac{1}
{2} \big( Q_j^{\vphantom{\ast}} + Q_j^\ast \big) =: \mathfrak{Re}
\, Q_j$ and $\frac{1}{2{\rm i}} \big( Q_j^{\vphantom{ \ast}} -
Q_j^\ast \big) =: \mathfrak{Im} \, Q_j$, you have $|F_M| = {\rm
e}^{- \frac{1}{4}(f_1 + f_2 + f_3)}$ with
\begin{eqnarray*}
    &&f_1(Q) = \sum_{ij} w_{ij} {\rm Tr} \, (s \, \mathfrak{Im} \,
    Q_i + z) (s \, \mathfrak{Im} \, Q_j + z) + \mbox{c.c.} \;, \\
    &&f_2(Q) = - 2 \sum_{ij} w_{ij} {\rm Tr} \, (s \, \mathfrak{Re}
    \, Q_i) ( s \, \mathfrak{Re} \, Q_j) \;, \\ &&f_3(Q) = 4 \sum_i {
    \rm Tr} \, \left( M_i + s \, \mathfrak{Im}\, z \,\,
    {\textstyle{\sum_j w_{ij}}} \right) \mathfrak{Re} \, Q_i \;.
\end{eqnarray*}
These expressions suggest making the following choice of
integration domain for $Q_i$ ($i = 1, \ldots, |\Lambda|$).  Pick
some real constant $\lambda > 0$ and put
\begin{displaymath}
    \mathfrak{Re} \, Q_i = \lambda \, T_i^{\vphantom{\ast}}
    T_i^{\ast} \;, \quad \mathfrak{Im} \, Q_i = P_i :=
    \begin{pmatrix} P_i^+ &0\\ 0 &P_i^- \end{pmatrix} \;,
\end{displaymath}
with $T_i \in {\rm U}(p,q)$, $P_i^+ \in {\rm Herm} (\mathbb{
C}^p)$, $P_i^- \in {\rm Herm}(\mathbb{C}^q)$.  The set of matrices
$Q_i$ so defined is referred to as the Sch\"afer-Wegner domain
$X_{\lambda}^{p,q}$.  The range of the field $Q = (Q_1, \ldots,
Q_{ |\Lambda|})$ is the direct product ${\mathcal X} := \big(
X_\lambda^ {p,q} \big)^{|\Lambda|}$.

To show that this is a good choice of domain, we first show
convergence of the integral $\int_{\mathcal X} F_M(Q) DQ$.  First
of all, the matrices $P_i$ commute with $s$, so
\begin{displaymath}
    f_1(Q)\Big|_{\mathcal X} = 2 \, \mathfrak{Re} \sum_{ij}
    w_{ij} \, {\rm Tr} \, (P_i + s z) (P_j + s z) \;.
\end{displaymath}
Since the coefficients $w_{ij}$ are positive as a quadratic form,
this expression is convex (with a positive Hessian) in the
Hermitian matrices $P_i$.  Second, the function
\begin{displaymath}
   f_2(Q)\Big|_{\mathcal X} = - 2\, \lambda^2 \sum_{ij} w_{ij} \,
   {\rm Tr} \, \left( T_i^{\vphantom{\ast}} T_i^\ast \right)^{-1}
   T_j^{\vphantom{\ast}} T_j^\ast
\end{displaymath}
is bounded from below by the constant $- 2\lambda^2 n \sum_i
w_{ii}$. This holds true because $w_{ij}$ is negative for $i \not=
j$, and because $T_i^{\vphantom{ \ast}} T_i^\ast > 0$ and the
trace of a product of two positive Hermitian matrices is always
positive. Third, 
\begin{displaymath}
    f_3(Q)\Big|_{\mathcal X} = 4 \, \lambda \sum_i {\rm Tr} \, \left(
    M_i + s \, \mathfrak{Im}\, z \,\, {\textstyle{\sum_j
    w_{ij}}} \right) T_i^{\vphantom{\ast}} T_i^\ast
\end{displaymath}
is positive, as $( \ldots )$ is positive Hermitian. As long as
$s\, \mathfrak{Im}\, z > 0$ the function $f_3$ goes to infinity
for all possible directions of taking the $T_i$ to infinity on
${\rm U}(p,q)$.

Thus when the matrices $Q_i$ are taken to vary on the
Sch\"afer-Wegner domain $X_\lambda^{p,q}$, the absolute value
$|F_M| = {\rm e}^{- \frac{1}{4} (f_1+f_2+f_3)}$ decays rapidly at
infinity.
%
%
This establishes the convergence of $\int_{\mathcal X} F_M(Q)DQ$.

Next, let us count dimensions.  The mapping $T \mapsto T T^\ast$
for $T \in {\rm U}(p,q) =: G$ is invariant under right
multiplication of $T$ by elements of the unitary subgroup $H :=
{\rm U} (p) \times {\rm U}(q)$ -- it is called the {\it Cartan
embedding} of $G/H$ into $G$.  The real manifold $G/H$ has
dimension $2pq$ and so does its image under the Cartan embedding.
Augmenting this by the dimension of ${\rm Herm}(\mathbb{C}^p)$ and
${\rm Herm} (\mathbb{ C}^q)$ (from $P_i$), one gets ${\rm dim}\,
X_\lambda^{p,q} = 2pq + p^2 + q^2 = (p + q)^2 = n^2$, which is as
it should be.

Finally, why can one shift variables and do the Gaussian integral
over $Q$ (with translation-invariant $DQ$) by completing the
square? This question is legitimate as the Sch\"afer-Wegner domain
$X_\lambda^{p,q}$ lacks invariance under the required shift, which
is $Q_i \mapsto Q_i - {\rm i}sz + \sum_j J_{ij} \, s M_j s$.

To complete the square in (\ref{funcFMQ}), introduce a parameter
$t \in [0,1[$ and consider the family of shifts
\begin{displaymath}
    Q_i \mapsto Q_i + t \big( - {\rm i} s z + {\textstyle{ \sum_j}}
    \, J_{ij} \, s M_j s \big) \;.
\end{displaymath}
For fixed $t$, this shift takes ${\mathcal X} = \big( X_\lambda^{
p,q} \big)^{| \Lambda|}$ into another domain, ${\mathcal X}(t)$.
Inspection shows that the function (\ref{funcFMQ}) still decays
rapidly (uniformly in the $M_i$) on ${\mathcal X}(t)$, as long as
$t < 1$. Without changing the integral one can add pieces to
${\mathcal X}(t)$ (for $t < 1$) at infinity to arrange for the
chain ${\mathcal X} - {\mathcal X}(t)$ to be a cycle. Because
${\mathcal X}(t)$ is homotopic to ${\mathcal X}(0) = {\mathcal
X}$, this cycle is a boundary: there exists a manifold ${\mathcal
Y}(t)$ of dimension ${\rm dim}\, {\mathcal X} + 1$ such that
$\partial {\mathcal Y}(t) = {\mathcal X} - {\mathcal X}(t)$.
Viewed as a holomorphic differential form of degree $(n^2
|\Lambda|, 0)$ in the complex space ${\rm End} (\mathbb{C}^n)^{
|\Lambda|}$, the integrand $\omega := F_M(Q) DQ$ is closed (i.e.
${\rm d}\omega = 0$). Therefore, by Stokes' theorem,
\begin{displaymath}
    \int_{\mathcal X} \omega - \int_{{\mathcal X}(t)} \omega =
    \int_{ \partial {\mathcal Y}(t)} \omega = \int_{{\mathcal Y}(t)}
    {\rm d}\omega = 0 \;,
\end{displaymath}
which proves $\int_{{\mathcal X}(t)} F_M(Q) DQ = \int_{\mathcal X}
F_M(Q) DQ$, independent of $t$. (This argument does \emph{not} go
through for the non-rigorous choice $sQ_i := T_i P_i T_i^{-1}$
usually made!)

In the limit $t \to 1$, one encounters the expression
\begin{eqnarray*}
    &&\int_{{\mathcal X}(1)} F_M(Q) DQ = \int_{\mathcal X}
    d\nu_{n,J}({\rm i}Q) \\ &&\hspace{2.6cm}
    \times \, {\rm e}^{- \frac{1}{2} \sum_{ij} J_{ij} {\rm Tr}
    \, ( s M_i s M_j ) + {\rm i} \sum_k {\rm Tr} \, (s z M_k)}
\end{eqnarray*}
with $d\nu_{n,J}$ as in (\ref{measure}).  The normalization
integral over ${\mathcal X}$ is \emph{defined} by taking the
Hermitian matrices $P_i$ to be the \emph{inner} variables of
integration. The outer integrals over the $T_i$ then demonstrably
exist, and one can fix the (otherwise arbitrary) normalization of
$DQ$ by setting $\int_{\mathcal X} d\nu_{n,J}({\rm i}Q) = 1$.
%
%
Making that choice, and comparing with (\ref{difficult}), one has
proved
\begin{displaymath}
\Omega_{n,N}^{\rm bos} = \int_{\varphi, \bar\varphi} \left(
\int_{\mathcal X} F_{(M_i)_{ \alpha\beta} = (\bar\varphi_\alpha,
\Pi_i \varphi_\beta)} (Q) DQ \right) \;.
\end{displaymath}

The final step is to change the order of integration over the $Q$-
and $\varphi$-variables, which is permitted since the $Q$-integral
converges uniformly in $\varphi$.
%
%
Doing the Gaussian $\varphi$-integral and shifting $Q_k \to Q_k -
{\rm i} s \, z$, one arrives at the Sch\"afer-Wegner formula for
$\Omega_{n,N}^{\rm bos}$:
\begin{eqnarray}
    &&\Omega_{n,N}^{\rm bos}(z,w^{-1}) = \int_{\mathcal X} {\rm
    e}^{\frac{1}{2} \sum_{ij} w_{ij} {\rm Tr} \, ( s Q_i s Q_j)}
    \label{wegner}\\ &&\hspace{2.6cm} \times \, {\rm e}^{- N
    \sum_k {\rm Tr}\, \ln ( Q_k - {\rm i}s z )} DQ \;, \nonumber
\end{eqnarray}
which is a rigorous version of the naive formula (\ref{naive}).
Compared to Fyodorov's formula, it has the disadvantage of not
being manifestly invariant under global hyperbolic transformations
$Q_i \mapsto T Q_i T^\ast$ (the integration domain ${\mathcal X}$
isn't invariant). Its best feature is that it does apply to the
case of band random matrices with one orbital per site ($N = 1$).
%

\section{Supersymmetric variant}

We are now in a position to tackle the problem of averaging ratios
of determinants. For concreteness, we shall discuss the case where
the number of determinants is two for both the numerator and the
denominator, which is what is needed for the calculation of the
function $G_{ij}^{(2)}(z_1,z_2)$ defined in equation
(\ref{twopoint}). We will consider the case of relevance for the
electrical conductivity: $z_1 = E + {\rm i}\epsilon$, $z_2 = E -
{\rm i}\epsilon$, with $E \in \mathbb{R}$ and $\epsilon > 0$.

A $Q$-integral formula for $G_{ij}^{(2)}(z_1 , z_2)$ can be
derived by combining the fermionic method for
\begin{displaymath}
    \big\langle {\rm Det}(z_1 - H) \, {\rm Det}(z_2 - H +
    t_2 E_{ij}^{ab}) \big\rangle
\end{displaymath}
with the Sch\"afer-Wegner bosonic formalism for
\begin{displaymath}
    \big\langle {\rm Det}^{-1}(z_1 - H - t_1 E_{ji}^{ba}) \,
    {\rm Det}^{-1} (z_2 - H) \big\rangle \;,
\end{displaymath}
and eventually differentiating with respect to $t_1, t_2$ at $t_1
= t_2 = 0$ and summing over $a,b$; see Section \ref{sec:gfd}. All
steps are formally the same as before, but with traces and
determinants replaced by their supersymmetric ana\-logs. Having
given a great many technical details in Sections \ref{sec:ferm}
and \ref{sec:bos}, we now just present the final formula along
with the necessary definitions and some indication of what are the
new elements involved in the proof.

Let $Q_{\rm BB}$, $Q_{\rm FF}$, $Q_{\rm BF}$ and $Q_{\rm FB}$ each
stand for a $2 \times 2$ matrix.  If the first two matrices have
commuting entries and the last two anti-commuting ones, they
combine to a $4 \times 4$ {\it supermatrix}:
\begin{displaymath}
    Q = \begin{pmatrix} Q_{\rm BB} &Q_{\rm BF}\\
    Q_{\rm FB} &Q_{\rm FF} \end{pmatrix} \;.
\end{displaymath}
Relevant operations on supermatrices are the supertrace,
\begin{displaymath}
    {\rm STr}\, Q = {\rm Tr}\, Q_{\rm BB} - {\rm Tr}\, Q_{\rm FF}
    \;,
\end{displaymath}
and the superdeterminant,
\begin{displaymath}
    {\rm SDet}\, Q = \frac{ {\rm Det}(Q_{\rm BB}) }{ {\rm Det}(
    Q_{\rm FF} - Q_{\rm FB} {Q_{\rm BB}}^{-1} Q_{\rm BF} ) } \;.
\end{displaymath}
These are related by the identity ${\rm SDet} = \exp \circ {\rm
STr} \circ \ln$ whenever the superdeterminant exists and is
nonzero.

In the process of applying the method described earlier, a
supermatrix $Q_i$ gets introduced at every site $i$ of the lattice
$\Lambda$. The domain of integration for each of the matrix blocks
$(Q_i)_{\rm BB}$ ($i = 1, \ldots, |\Lambda|$) is taken to be the
Sch\"afer-Wegner domain $X_\lambda^{1,1}$ (with some choice of
$\lambda > 0$); the integration domain for each of the $(Q_i)_{\rm
FF}$ is the space of Hermitian $2 \times 2$ matrices, as before.

Let $E_{\rm BB}^{11}$ be the $4\times 4$ (super)matrix with unit
entry in the upper left corner and zeroes elsewhere; similarly,
$E_{\rm FF}^{22}$ has unity in the lower right corner and zeroes
elsewhere. Putting $s = {\rm diag}(1,-1,1,1)$ and $z = {\rm diag}(
z_1, z_2, z_1, z_2)$, the supersymmetric $Q$-integral formula for
the generating function of $G_{ij}^{(2)}$ -- obtained by combining
the Sch\"afer-Wegner bosonic method with the fermionic variant --
is written as
\begin{eqnarray}
    &&\left\langle \frac{{\rm Det}(z_1 - H) \, {\rm Det}(z_2 - H
    + t_2 E_{ij}^{ab})} {{\rm Det}(z_1 - H - t_1 E_{ji}^{ba})
    \, {\rm Det}( z_2 - H)} \right\rangle \label{susygen} \\
    &&= \int DQ \, \, {\rm e}^{\frac{1}{2} \sum_{kl} w_{kl}\,
    {\rm STr} \, ( s Q_k s Q_l)} \nonumber \\
    &&\times {\rm e}^{- {\rm STr} \, \ln \big( \sum_{r,c}
    (Q_r - {\rm i}s z) \otimes E_{rr}^{cc}
    + {\rm i}\, t_1 E_{\rm BB}^{11} \otimes E_{ji}^{ba}
    - {\rm i}\, t_2 E_{\rm FF}^{22} \otimes E_{ij}^{ab} \big) },
    \nonumber
\end{eqnarray}
where the second supertrace includes a sum over sites and
orbitals, and on setting $t_1 = t_2 = 0$ becomes
\begin{displaymath}
    {\rm e}^{-N \sum_r {\rm STr}\, \ln (Q_r - {\rm i}sz)} =
    {\textstyle{\prod_r}} {\rm SDet}^{-N} (Q_r - {\rm i}sz) \;.
\end{displaymath}
The superintegral `measure' $DQ = \prod_r D Q_r$ is the {\it flat
Berezin form}, i.e.~the product of differentials for all the
commuting matrix entries in $(Q_r)_{\rm BB}$ and $(Q_r)_{\rm FF}$,
times the product of derivatives for all the anti-commuting matrix
entries in $(Q_r)_{\rm BF}$ and $(Q_r)_{\rm FB}$.

To prove the formula (\ref{susygen}), two new tools are needed, a
brief account of which is as follows.

\subsection{Gaussian superintegrals}

There exists a supersymmetric generalization of the Gaussian
integration formulas given in Section \ref{sec:dgi}: if $A,D$
($B,C$) are linear operators or matrices with commuting
(resp.~anti-commuting) entries, and $\mathfrak{Re}\, A > 0$, one
has
\begin{displaymath}
    {\rm SDet}^{-1} \begin{pmatrix} A &B\\ C &D \end{pmatrix}
    = \int {\rm e}^{ - (\bar\varphi, A \varphi) - (\bar\varphi ,
    B \psi) - (\bar\psi, C \varphi) - (\bar\psi , D \psi)} \;.
\end{displaymath}
Verification of this formula is straightforward.  Using it, one
writes the last factor in (\ref{susygen}) as a Gaussian
superintegral over four vectors: $\varphi_1$, $\varphi_2$,
$\psi_1$, and $\psi_2$.  The integrand then becomes Gaussian in
the matrices $Q_r$.

\subsection{Shifting variables}

The next step in the proof is to do the `Gaussian' integral over
the supermatrices $Q_r$.  By definition, in a superintegral one
first carries out the Fermi integral, and afterwards the ordinary
integrations.  The Gaussian integral over the anti-commuting parts
$(Q_r)_{\rm BF}$ and $(Q_r)_{\rm FB}$ is readily done by
completing the square and shifting variables using the fact that
fermionic integration is differentiation:
\begin{displaymath}
\int d\xi\, f(\xi - \xi^\prime) = \frac{\partial}{\partial\xi}
f(\xi - \xi^\prime) = \int d\xi\, f(\xi) .
\end{displaymath}
Similarly, the Gaussian integral over the Hermitian matrices
$(Q_r)_{\rm FF}$ is done by completing the square and shifting.
The integral over $(Q_r)_{\rm BB}$, however, is \emph{not}
Gaussian, as the domain is not $\mathbb{R}^n$ but the
Sch\"afer-Wegner domain. Here, more advanced calculus is required:
these integrations are done by using a supersymmetric
change-of-variables theorem due to Berezin to make the necessary
shifts by {\it nilpotents}. (There is not enough space to describe
this here, so please consult Berezin's book.)
Without difficulty one finds the result to agree with the
left-hand side of Eq.~(\ref{susygen}), thereby establishing that
formula.

\section{Approximations}\label{sec:approx}

All manipulations so far have been exact and, in fact, rigorous
(or can be made so with little extra effort). Now we turn to a
sequence of approximations that have been used by physicists to
develop a quantitative understanding of weakly disordered quantum
dots, wires, films etc.  While physically satisfactory, not all of
these approximations are under full mathematical control. We will
briefly comment on their validity as we go along.

\subsection{Saddle-point manifold}

We continue to consider $G_{ij}^{(2)}(E +{\rm i}\varepsilon, E -
{\rm i}\varepsilon)$ and focus on $E = 0$ (the center of the
energy band) for simplicity. By varying the exponent on the
right-hand side of (\ref{susygen}) at $t_1 = t_2 = 0$, one gets
the following equation:
\begin{displaymath}
    \sum_j w_{ij} \, s \, Q_j \, s - N Q_i^{-1} = 0 \;,
\end{displaymath}
which is called the saddle-point equation.

Let us now assume {\it translational invariance}, $w_{ij} =
f(|i-j|)$. Then, if $\lambda = \sqrt{N / \sum_j w_{ij}}$ , the
saddle-point equation has $i$-independent solutions of the form
\begin{displaymath}
    Q_i = \lambda \begin{pmatrix} q_{\rm BB} &0 \\ 0 &q_{\rm FF}
    \end{pmatrix},
\end{displaymath}
where for $q_{\rm FF}$ there are {\it three} possibilities: two
isolated points $q_{\rm FF} = \pm {\bf 1}$ (unit matrix) coexist
with a manifold
\begin{equation}\label{sphere}
   q_{\rm FF} = \begin{pmatrix}
    \cos \theta_1 &\sin \theta_1 \, {\rm e}^{{\rm i}\phi_1} \\
    \sin \theta_1 \, {\rm e}^{-{\rm i}\phi_1} & -\cos\theta_1
    \end{pmatrix} \;,
\end{equation}
which is $2$-dimensional, whereas the solution space for $q_{\rm
BB}$ consists of a \emph{single} connected $2$-manifold:
\begin{equation}\label{hyperbol}
    q_{\rm BB} = \begin{pmatrix}
    \cosh \theta_0 &\sinh \theta_0 \, {\rm e}^{{\rm i}\phi_0} \\
    \sinh \theta_0 \, {\rm e}^{-{\rm i}\phi_0} & \cosh\theta_0
    \end{pmatrix} \;.
\end{equation}
The solutions $q_{\rm FF} = \pm {\bf 1}$ are usually discarded in
the physics literature.  (The argument is that they break
supersymmetry and therefore get suppressed by fermionic zero
modes. For the simpler case of the one-point function
(\ref{onepoint}) and in three space dimensions, such suppression
has recently been proved by Disertori, Pinson and Spencer.) Other
solutions for $q_{\rm BB}$ are ruled out by the requirement
$\mathfrak{Re}\, Q_i > 0$ for the Sch\"afer-Wegner domain.

The set of matrices (\ref{hyperbol}) and (\ref{sphere}) -- the
{\it saddle-point manifold} -- is diffeomorphic to the product of
a two-hyperboloid ${\rm H}^2$ with a two-sphere ${\rm S}^2$.
Moving along that manifold $M := {\rm H}^2 \times {\rm S}^2$
leaves the $Q$-field integrand (\ref{susygen}) unchanged (for $z_1
= z_2 = t_1 = t_2 = 0$).

One can actually anticipate the existence of such a manifold from
the symmetries at hand. These are most transparent in the starting
point of the formalism as given by the characteristic function
$\langle {\rm e}^{-{\rm i} {\rm Tr}\, H K} \rangle$ with
\begin{displaymath}
    K = \bar\varphi_1 \otimes \varphi_1 - \bar\varphi_2 \otimes
    \varphi_2 + \bar\psi_1 \otimes \psi_1 + \bar\psi_2 \otimes
    \psi_2 \;.
\end{displaymath}
The signs of this quadratic expression are what is encoded in the
signature matrix $s = {\rm diag}(1,-1,1,1)$ (recall that the first
two entries are forced by $\mathfrak{Im}\, z_1 > 0$ and
$\mathfrak{Im} \, z_2 < 0$). The quadratic form $K$ is invariant
under the product of two Lie groups: ${\rm U}(1,1)$ acting on the
$\varphi$'s, and ${\rm U}(2)$ acting on the $\psi$'s.  This
invariance gets transferred by the formalism to the $Q$-side; the
saddle-point manifold $M$ is in fact an {\it orbit} of the group
action of $G := {\rm U}(1,1) \times {\rm U}(2)$ on the $Q$-field.
In the language of physics, the degrees of freedom of $M$
correspond to the {\it Goldstone bosons} of a broken symmetry.

$K$ also has a number of supersymmetries, mixing $\varphi$'s with
$\psi$'s. At the infinitesimal level, these combine with the
generators of $G$ to give a {\it Lie superalgebra} of symmetries
$\mathfrak{g} := \mathfrak{u}(1,1|2)$.  One therefore expects some
kind of saddle-point {\it supermanifold}, say $\mathcal{M}$, on
the $Q$-side.

$\mathcal{M}$ can be constructed by extending the above solution
$q_0 := {\rm diag}(q_{\rm BB},q_{\rm FF})$ of the dimensionless
saddle-point equation $s q s = q^{-1}$ to the full $4 \times 4$
supermatrix space.  Putting $q = q_0 + q_1$ with $q_1 =
\begin{pmatrix} 0 &q_{\rm BF}\\ q_{\rm FB} &0 \end{pmatrix}$, and
linearizing in $q_1$, one gets
\begin{equation}\label{fibre}
    s \, q_1 \, s = - q_0^{-1} q_1^{\vphantom{-1}} q_0^{-1} \;.
\end{equation}
The solution space of this linear equation for $q_1$ has dimension
four for all $q_0 \in M$. Based on it, one expects four Goldstone
{\it fermions} to emerge along with the four Goldstone bosons of
$M$.

For the simple case under consideration, one can introduce local
coordinates and push the analysis to non-linear order, but things
get quickly out of hand (when done in this way) for more
challenging, higher-rank cases. Fortunately, there exists an
alternative, coordinate-independent approach, as the mathematical
object to be constructed is completely determined by symmetry!

\subsection{Riemannian symmetric superspace}

The linear equation (\ref{fibre}) associates with every point $x
\in M$ a four-dimensional vector space of solutions $V_x$.  As the
point $x$ moves on $M$ the vector spaces $V_x$ turn and twist;
thus they form what is called a {\it vector bundle} $V$ over $M$.
(The bundle at hand turns out to be non-trivial, i.e.~there exists
no global choice of coordinates for it.)

A {\it section} of $V$ is a smooth mapping $s : M \to V$ such that
$s(x) \in V_x$ for all $x \in M$.  The sections of $V$ are to be
multiplied in the exterior sense, as they represent anti-commuting
degrees of freedom; hence the proper object to consider is the
{\it exterior bundle}, $\wedge V$.

It is a beautiful fact that there exists a unique action of the
Lie superalgebra $\mathfrak{g}$ on the sections of $\wedge V$ by
first-order differential operators, or {\it derivations} for
short.  (Be advised however that this canonical $\mathfrak{
g}$-action is not well-known in physics or mathematics.)

The manifold $M$ is a {\it symmetric space}, i.e.~a Riemannian
manifold with $G$-invariant geometry.  Its metric tensor, $g$,
uniquely extends to a second-rank tensor field (still denoted by
$g$) which maps pairs of derivations of $\wedge V$ to sections of
$\wedge V$, and is invariant with respect to the $\mathfrak{g
}$-action. This collection of objects --- the symmetric space $M$,
the exterior bundle $\wedge V$ over it, the action of the Lie
superalgebra $\mathfrak{g}$ on the sections of $\wedge V$, and the
$\mathfrak{g}$-invariant second-rank tensor $g$ --- form what the
author calls a {\it Riemannian symmetric superspace},
$\mathcal{M}$.

\subsection{Non-linear sigma model}

According to the Landau-Ginzburg-Wilson paradigm of the theory of
pha\-se transitions, the large-scale physics of a statistical
mechanical system near criticality is expected to be controlled by
an effective field theory for the long-wave length excitations of
the {\it order parameter} of the system.

Wegner is credited for the profound insight that the LGW paradigm
applies to the random-matrix situation at hand, with the role of
the order parameter being taken by the matrix $Q$.  He argued that
transport observables (such as the electrical conductivity) are
governed by slow spatial variations of the $Q$-field inside the
saddle-point manifold. Efetov skilfully implemented this insight
in a supersymmetric variant of Wegner's method.

While the direct construction of the effective continuum field
theory by gradient expansion of (\ref{susygen}) is not an entirely
easy task, the outcome of the calculation is pre-determined by
symmetry.  On general grounds, the effective field theory has to
be a {\it non-linear sigma model} for the Goldstone bosons and
fermions of $\mathcal{M}$: if $\{ \phi^A \}$ are local coordinates
for the bundle $V$ with metric $g_{AB}(\phi) = g(\partial/
\partial\phi^A , \partial/ \partial\phi^B)$, the action
functional is
\begin{displaymath}
    S = \sigma \int d^d x \,\, \partial_\mu \phi^A \, g_{AB}(\phi)
    \, \partial_\mu \phi^B \;.
\end{displaymath}
The coupling parameter $\sigma$ has the physical meaning of bare
(i.e.~unrenormalized) conductivity.  In the present model $\sigma
= N W^2 a^{2-d}$, where $W$ is essentially the width of the band
random matrix in units of the lattice spacing $a$ (the
short-distance cutoff of the continuum field theory). $S$ is the
effective action in the limit $z_1 = z_2$. For a finite frequency
$\omega = z_1 - z_2$, a symmetry-breaking term of the form ${\rm
i} \omega \nu \int d^d x\, f(\phi)$, where $\nu = N (\pi
\lambda)^{-1} a^{-d}$ is the local density of states, has to be
added to $S$.
%

By perturbative {\it renormalization group analysis}, i.e. by
integrating out the rapid field fluctuations, one finds for $d =
2$ that $\sigma$ decreases on increasing the cutoff $a$. This
property is referred to as {\it asymptotic freedom} in field
theory. On its basis one expects exponentially decaying
correlations, and hence localization of all states, in two
dimensions.  However, a mathematical proof of this conjecture is
not currently available.

In three dimensions and for a sufficiently large bare
conductivity, the renormalization flow goes toward the {\it
metallic} fixed point ($\sigma \to \infty$), where $G$-symmetry is
broken spontaneously. A rigorous proof of this important
conjecture (existence of disordered metals in three space
dimensions) is not available either.

\subsection{Zero-mode approximation}

For a system in a box of linear size $L$, the cost of exciting
fluctuations in the sigma model field is estimated as the {\it
Thouless energy} $E_{\rm Th} = \sigma / \nu L^2$.  In the limit of
small frequency, $|\omega| \ll E_{\rm Th}$, the physical behavior
is dominated by the constant modes $\phi^A(x) = \phi^A$
(independent of $x$). By computing the integral over these modes,
Efetov found the energy-level correlations in the small-frequency
limit to be those of the Gaussian Unitary Ensemble.

\end{document}